\documentclass[aps,prl,showpacs,twocolumn,superscriptaddress]{revtex4}
\usepackage{bm,color}
\usepackage{graphicx}
\usepackage{amsmath, amssymb}

\begin{document}
\title{Proposed Negative Thermal Expansion in Honeycomb-Lattice Antiferromagnets}
\author{Yuto Uwabo}
\author{Masahito Mochizuki}
\affiliation{Department of Applied Physics, Waseda University, Okubo, Shinjuku-ku, Tokyo 169-8555, Japan}
\begin{abstract}
We theoretically propose possible magnetism-induced negative thermal expansion in honeycomb-lattice antiferromagnets with edge-sharing networks of $MX_6$ octahedra where $M$ and $X$ are transition-metal and ligand ions, respectively. In this crystal structure, the nearest-neighbor exchange interaction is composed of two competing contributions, i.e., the antiferromagnetic contribution from a direct 180$^\circ$ $M$-$M$ bond and the ferromagnetic contribution from 90$^\circ$ $M$-$X$-$M$ bonds, amplitudes of which have different bond-length dependence. Numerical analysis of the spin-lattice model of the honeycomb-lattice antiferromagnets demonstrates that the negative thermal expansion can occur when the system enters the antiferromagnetic phase with lowering temperature so as to maximize the energy gain associated with the bond-length dependent antiferromagnetic exchange interaction. The present work provides a guiding principle for searching new materials and eventually contributes to diversify the family of materials that host the negative thermal expansion originating from the spin-lattice coupling on the honeycomb lattices or related crystal structures.
\end{abstract}
\maketitle

\section{Introduction}
Ordinary materials expand (contract) in volume as temperature increases (decreases)~\cite{ZimanTB}. However, there are several rare examples of materials that exhibit an opposite behavior, i.e., expand (contract) upon cooling (heating)~\cite{ChuCN87,Sleight98,Barrera05,Takenaka12,Takenaka14,ChenJ15,Takenaka18}. This unusual thermal volume effect is called negative thermal expansion. Research on such materials has more than 120-year long history since the discovery of invar alloys in 1897~\cite{Guillaume1897}. The thermally induced changes of crystal volumes often cause serious problems in modern technologies. For highly developed precision machineries and optical devices, the changes of volumes and/or lengths can fatally spoil their functions. Combinations of materials with different rates of thermal expansion often suffer the peeling and dropping off. Therefore, managements of the thermal volume effects become more and more important nowadays, which have attracted a great deal of interest from viewpoints of both fundamental science and industrial technology. Because the crystal-volume change due to temperature variation can be suppressed by using a composite of normal material and negative-thermal-expansion material, the research of negative thermal expansion is now rapidly developed along with the growth of the demands from society.

However, physical mechanisms of the negative thermal expansion have not been fully clarified yet in spite of the long history of research. This is because this phenomenon is often caused by complex couplings among multi degrees of freedom in materials, e.g., spins, charges and orbitals of electrons and crystal lattices. Thus, search for new materials have long been depending on experiences and intuitions of researchers. Under this circumstance, an important guiding principle for the material search has been proposed recently from a theoretical study~\cite{Kobayashi19}. This study has revealed that competing two opposite contributions to the nearest-neighbor antiferromagnetic exchange coupling and their different bond-length dependence can be a source of the negative thermal expansion by taking inverse perovskite manganese nitrides Mn$_3$$A$N ($A$=Zn, Ga, Cu$_{1-x}$Ge$_x$ etc) as an important class of materials exhibiting a large crystal-volume expansion triggered by the magnetic phase transition~\cite{Takenaka12,Takenaka14,Bouchaud68,Fruchart71,Takenaka05,Takenaka06,Takenaka08,Hamada11,HuangR08,SunZH09,SunY07,SunY10,SongXY11}. The competing contributions are antiferromagnetic one from the direct $d$-$d$ electron hopping between the adjacent $M$ ions and ferromagnetic one from the indirect $d$-$d$ electron hopping mediated by an in-between ligand ion on the 90$^\circ$ bond. From this work, we have learned that antiferromagnets whose nearest-neighbor exchange interaction is composed of two opposite contributions potentially host large magnetism-driven negative thermal expansion.

\begin{figure}[tb]
\includegraphics[scale=0.25]{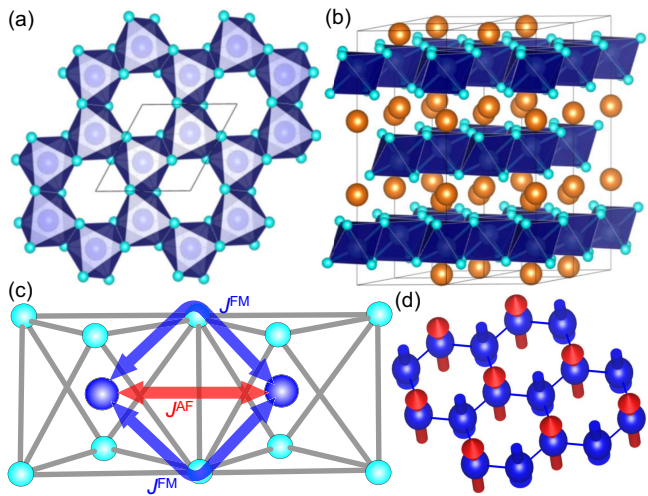}
\caption{(Color online) (a) Honeycomb lattice composed of edge-sharing $MX_6$ octahedra where $M$ and $X$ are transition-metal and ligand ions, respectively. (b) Ilmenite-type crystal structure as an example of the structures containing honeycomb lattices. (c) Nearest-neighbor exchange interaction composed of two opposite contributions, i.e., the antiferromagnetic contribution from the direct $M$-$M$ bond and the ferromagnetic contribution from the 90$^\circ$ $M$-$X$-$M$ bonds. (d) Antiferromagnetic spin order on the honeycomb lattice. The software VESTA was used for drawing the crystal and magnetic structures~\cite{VESTA}.}
\label{Fig01}
\end{figure}

In this paper, we expand this idea to another class of materials, i.e., honeycomb-lattice magnets. We construct a theoretical model to describe the spin-lattice coupled system in transition-metal compounds whose crystal structure contains honeycomb lattices composed of edge-sharing $MX_6$ octahedra where $M$ and $X$ are transition-metal and ligand ions, respectively [Figs.~\ref{Fig01}(a) and (b)]. On the basis of theoretical analyses of this model, we propose that large negative thermal expansion through this mechanism can be observed when certain conditions are satisfied. The exchange spin coupling between adjacent edge-sharing $MX_6$ octahedra has both the antiferromagnetic contribution from the direct 180$^\circ$ $M$-$M$ bond and the ferromagnetic contribution from 90$^\circ$ $M$-$X$-$M$ bonds [Fig.~\ref{Fig01}(c)], amplitudes of which have different bond-length dependencies. Note that the exchange interactions in ordinary materials without such competing contributions are usually weakened when the crystal volume expands because the bond elongation necessarily reduces the hybridization between the $M$-ion $d$ orbitals. On the contrary, the nearest-neighbor exchange interactions in the honeycomb-lattice antiferromagnets can be strengthened upon the crystal-volume expansion because the ferromagnetic contribution is strongly suppressed rather than the antiferromagnetic contribution when the bond is elongated. Consequently, the honeycomb-lattice antiferromagnets can exhibit a crystal-volume expansion when the system enters the antiferromagnetically ordered phase [Fig.~\ref{Fig01}(d)] with decreasing temperature so as to maximize the energy gain associated with the nearest-neighbor antiferromagnetic exchange interactions. We derive conditions under which the magnetism-driven negative thermal expansion occurs in the honeycomb-lattice antiferromagnets and demonstrate its validity by analyzing the constructed theoretical model using the Monte-Carlo technique. The present work provides an important guiding principle for the material search and will diversify the family of materials hosting the negative thermal expansion. There are a huge number of honeycomb lattice transition-metal compounds which exhibit antiferromagnetic~\cite{Shirane1959,Chemberland1970,Tsuzuki1974,Regnault1980,Rogado2002,Tsirlin2010,Pascual2012,LeeS2012,LeeS2014,YanYJ2012,Iakovleva2019,Rani2013,Seibel2013,Tang2014,Rao2014,Rani2015,Itoh2015,McNally2015,KimSW2016,Bera2017,Sugawara2017,Haraguchi2018,Haraguchi2019,Nalbandyan2019,Tursun2019,NiM2020,Sala2021,TangYS2021,Ishii2021,Smirnova2009} and ferrimagnetic~\cite{Nakayama2011,WangY2015,KimSW2016b,WangW2021} phases or related crystal structures~\cite{ZhangB2012,Cao2015,Zhang2015,Kumada2015,Zhou2016,Cho2017,Otsuka2018}. These compounds are candidate hosts of the magnetism-driven negative thermal expansion based on the proposed mechanism.

\section{Spin-Lattice Model}
We start with the following theoretical model to describe the spin-lattice coupling on a honeycomb lattice:
\begin{eqnarray}
{\mathcal H}&=& 
\sum_{\langle i,j\rangle}J(\delta_{ij})\textbf{S}_{i} \cdot \textbf{S}_{j}
-A\sum_{i}S^{2}_{iz} \nonumber \\ 
& &\hspace{2cm}
+K_1 \sum_{\langle i,j\rangle}\delta^{2}_{ij}
-K_2 \sum_{\langle i,j\rangle}\delta^{3}_{ij}.
\label{eq:model}
\end{eqnarray}
The first term describes the antiferromagnetic exchange interactions between the nearest-neighbor spins on the adjacent sites. Here $\textbf{S}_i$ denotes the classical spin vector at site $i$, and $\langle i,j\rangle$ denotes the adjacent site pair. The coupling constant $J(\delta_{ij})$ depends on the length $\ell_{ij}=\ell_0 + \Delta\ell_{ij}$ of the bond connecting sites $i$ and $j$ where $\ell_0$ is the standard lattice constant and $\Delta\ell_{ij}$ is a deviation of the bond length from $\ell_0$. The normalized deviation of the bond length is defined as $\delta_{ij} \equiv \Delta\ell_{ij}/\ell_0$. The second term describes the easy-axis magnetic anisotropy ($A>0$), which is known to exist in real honeycomb-lattice antiferromagnets, e.g., MgMnO$_3$~\cite{Haraguchi2019}. The third and fourth terms describe the harmonic and higher-harmonic components of elastic energies, respectively. We analyze this classical spin-lattice model using the replica-exchange Monte Carlo technique where the heat bath method is adopted to update the spins $\textbf{S}_i$ and the bond-length variation $\delta_{ij}$. Here we mention that a theoretical analysis based on a localized spin picture was successfully applied to the anomalous thermal volume effect in invar alloys previously~\cite{Hausch73}.

The nearest-neighbor exchange interaction is composed of opposite two contributions. One is a contribution from direct electron hopping on the 180$^\circ$ $M$-$M$ bond, which is expected to be antiferromagnetic according to the Kanamori-Goodenough rule~\cite{Kanamori59,Kanamori60,Goodenough55,Goodenough58}. Another contribution originates from indirect electron hopping between adjacent $M$ ions mediated by the in-between $X$ ion on the 90$^\circ$ $M$-$X$-$M$ bonds, which is expected to be ferromagnetic according to the Kanamori-Goodenough rule~\cite{Kanamori59,Kanamori60,Goodenough55,Goodenough58}. The former contribution $J^{\rm AF}$ is governed by the second-order perturbation processes with respect to the transfer integral $t_{dd}$ between $d$ orbitals on the adjacent $M$ ions, whereas the latter contribution $J^{\rm FM}$ is governed by the fourth-order perturbation processes with respect to the transfer integral $t_{pd}$ between $p$ and $d$ orbitals on the adjacent $X$ and $M$ ions. More specifically, the perturbation-expansion calculations give their explicit formulae as,
\begin{eqnarray}
J^{\rm AF}=\frac{4t_{dd}^2}{U}, \quad
J^{\rm FM}=-\frac{4t_{dp}^4 J_{\rm H}}{\Delta^2U^2}.
\end{eqnarray}
Here $U$ and $J_{{\rm H}}$ are strengths of the repulsive Coulomb interaction and the Hund's-rule coupling, respectively, whereas $\Delta$ is the charge-transfer energy between the $d$ and $p$ orbitals. Importantly, these two contributions have different bond-length dependence because the transfer integrals $t_{dd}$ and $t_{pd}$ have inherent distance dependence determined by radial parts of the wavefunctions of atomic orbitals~\cite{HarrisonTB}. Specifically, the transfer integral $t_{dd}$ between adjacent $d$ orbitals is proportional to $\ell^{-5}$, whereas $t_{pd}$ between adjacent $p$ and $d$ orbitals is proportional to $\ell^{-7/2}$ where $\ell$ is the distance between the atoms. 

Now we discuss the bond-length dependence of the coupling coefficient $J(\delta)$. Because the relationships $J^{\rm AF} \propto t_{dd}^2$ and $t_{dd} \propto \ell^{-5}$ hold, the bond-length dependence of $J^{\rm AF}$ is given by,
\begin{eqnarray}
J^{\rm AF}(\delta)=J_{\rm AF}(1+\delta)^{-10} \approx J_{\rm AF}(1-10\delta),
\end{eqnarray}
where $\ell=\ell_0(1+\delta)$. The coefficient $J_{\rm AF}$ corresponds to the antiferromagnetic contribution when the bond length is $\ell_0$. On the other hand, because the relationships $J^{\rm FM} \propto t_{pd}^4$ and $t_{pd} \propto \ell^{-7/2}$ hold, the bond-length dependence of $J^{\rm FM}$ is given by,
\begin{eqnarray}
J^{\rm FM}(\delta)=J_{\rm FM}(1+\delta)^{-14} \approx J_{\rm FM}(1-14\delta),
\end{eqnarray}
The coefficient $J_{\rm FM}$ corresponds to the ferromagnetic contribution when $\ell=\ell_0$. Here the standard lattice constant $\ell_0$ is defined as the bond length at $T$=0 in the absence of the spin-lattice coupling where the exchange interaction is independent of $\delta$.

The bond-length dependence of the nearest-neighbor exchange interaction $J$ is given by a sum of these two contributions as $J=J^{\rm AF}+2J^{\rm FM}$. Here the factor 2 appears because there are two $M$-$X$-$M$ paths for the neighbored edge-sharing $MX_6$ octahedra. Eventually we obtain
\begin{eqnarray}
J(\delta)
&=&J^{\rm AF}(\delta)+2J^{\rm FM}(\delta)
\nonumber \\
&\sim&J_{\rm AF}(1-10\delta)-2\left| J_{\rm FM} \right| (1-14\delta)
\nonumber \\
&=&(28\left| J_{\rm FM} \right|-10J_{\rm AF})\delta+(J_{\rm AF}-2\left| J_{\rm FM} \right|)
\label{eq:nnexch}
\end{eqnarray}\par

\section{Results}
We substitute the expression of $J(\delta)$ in Eq.~(\ref{eq:nnexch}) into the Hamiltonian in Eq.~(\ref{eq:model}) and analyze this Hamiltonian using the replica-exchange Monte-Carlo technique. In the Monte-Carlo simulations, we update not only the spin vectors $\bm S_i$ but also the bond-length deviations $\delta_{ij}$ at finite temperatures using the Metropolis algorithm to produce thermal equilibrium states and take statistical samplings to calculate thermal averages of several physical quantities. We adopt a honeycomb-lattice system with $20^2$ unit cells for the simulations where one unit cell contains two spin sites.

We first calculate several physical quantities to study the phase transitions.
\begin{figure}[tb]
\includegraphics[scale=1]{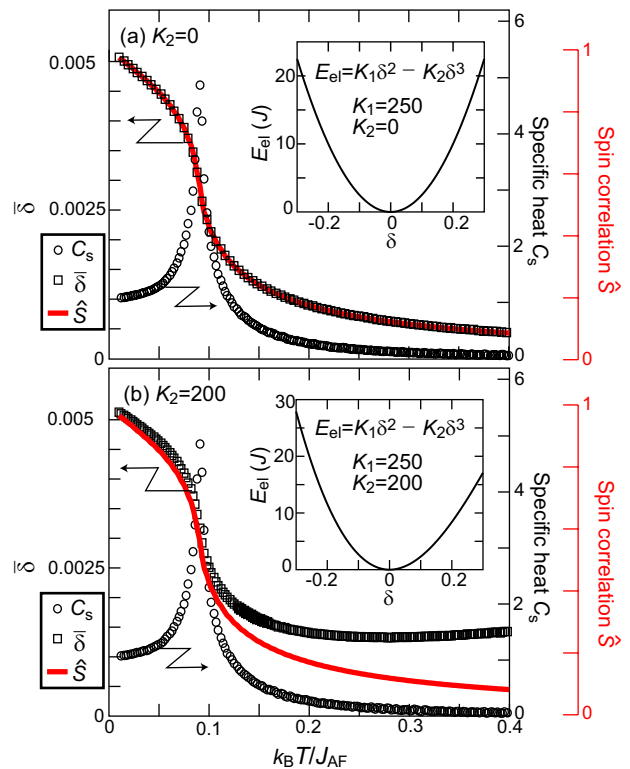}
\caption{(Color online) (a) Calculated temperature profiles of specific heat $C_{\rm s}$, averaged difference of normalized bond length $\bar{\delta}=(2/3N)\sum_{<i,j>}\langle \delta_{ij} \rangle$, and spin correlation $\hat{S}=(-1/N_{\rm pair})\sum_{<i,j>} \langle \bm S_{i}\cdot\bm S_{j} \rangle$ when the lattice elastic energy contains only the harmonic component with $K_1=250$ and $K_2=0$. (b) Those when the lattice elastic energy is anharmonic with $K_1=250$ and $K_2=200$. Insets in (a) and (b) show harmonic and anharmonic behaviors of the lattice elastic energy $E_{\rm el}$, respectively.}
\label{Fig02}
\end{figure}
Figure~\ref{Fig02}(a) shows calculated temperature profiles of specific heat $C_{\rm s}$, averaged difference of normalized bond length $\overline{\delta}$, and spin correlation $\hat{S}$ when the lattice elastic energy is a harmonic one with $K_1=250$ and $K_2=0$. We set $J_{\rm FM}=-0.45J_{\rm AF}$ for the calculations. The quantities $\hat{S}$ and $\overline{\delta}$ are defined by,
\begin{eqnarray}
\hat{S}=
-\frac{1}{N_{\rm pair}}\sum_{\left< i,j \right>}\left<{\bm S}_{i}\cdot {\bm S}_{j}\right>,
\quad
\bar{\delta}=
\frac{2}{3N}\sum_{<i,j>}\langle \delta_{i,j} \rangle,
\end{eqnarray}
where $N$ and $N_{\rm pair}$ are the numbers of lattice sites and the nearest-neighbor site pairs, respectively, while $\sum_{<i,j>}$ denotes the summation over the nearest-neighbor site pairs. Noticeably, the specific heat $C_{\rm s}$ has a sharp peak at $k_{\rm B}T/J_{\rm AF}\sim 1$, which indicates that a magnetic phase transition occurs at $T_{\rm c}\sim J_{\rm AF}/k_{\rm B}$. The spin correlation $\hat{S}$ is used to identify the type of magnetic order. This quantity takes a positive (negative) value in the presence of antiferromagnetic (ferromagnetic) correlation. Note that $\hat{S}=+1$ corresponds to a perfect staggered spin configuration, whereas $\hat{S}=-1$ corresponds to a fully polarized spin configuration. We find that the quantity $\hat{S}$ exhibits an abrupt increase below $T_{\rm c}$ and takes an almost saturated value of $\hat{S}=+1$ at the lowest temperature, indicating that the antiferromagnetic order emerges below $T_{\rm c}$.

We find that $\overline{\delta}$ increases as temperature decreases, which indicates the occurrence of crystal-volume expansion. Importantly, the increase of $\overline{\delta}$ becomes abrupt below $T_{\rm c}$, and its profile perfectly coincides with the growth of antiferromagnetic spin correlation $\hat{S}$, indicating that the antiferromagnetic order on the honeycomb lattice indeed triggers the negative thermal expansion.

In Fig.~\ref{Fig02}(b), we present calculated temperature profiles of the same physical quantities when the lattice elastic energy contains an anharmonic component with finite $K_2$ as $K_1=250$ and $K_2=200$. We find that the behaviors of these quantities are not affected by the anharmonic component even quantitatively except for the behavior of $\bar{\delta}$ in the temperature regime above $T_{\rm c}$. When the elastic energies are harmonic with $K_2=0$, the bond length monotonically increases upon cooling [Fig.~\ref{Fig02}(a)]. On the contrary, when the elastic energy is anharmonic with $K_2=200$, the bond length decreases upon cooling (or increases upon heating) above $T_{\rm c}$ [Fig.~\ref{Fig02}(b)]. This behavior is nothing but the normal thermal expansion, i.e., the volume contraction (expansion) upon cooling (heating). The consideration of the anharmonic nature of elastic energy is necessary to reproduce the normal thermal expansion above $T_{\rm c}$~\cite{KittelTB}. We have, however, learned that the arguments on the magnetism-induced negative thermal expansion below $T_{\rm c}$ are not affected even if this anharmonicity is neglected. Therefore, we will discuss the results obtained for $K_2=0$ in the following.

It should be noted that lengths of neighboring bonds in a crystallographic network are mutually correlated, and thus the bond-length variations occur cooperatively in real materials. On the contrary, our model neglects this kind of bond cooperative effects and treats the bond-length deviations $\delta_{ij}$ as individual degrees of freedom localized at the bonds. Because of this approximate treatment, our calculations tend to overestimate the entropies, particularly those associated with the lattice degrees of freedom. We expect, however, that this overestimate cause only negligible influence and the results and conclusions of this work are never affected qualitatively. Indeed, we have also examined a hypothetical situation that all the $M$-$M$ bonds have the same length where the bond-length changes occur uniformly in the system. This situation is opposite to the above situation, for which the entropies are underestimated, and we have found that the temperature profiles of the physical quantities shown in Fig.~\ref{Fig02} are not altered even semi-quantitatively.

\begin{figure}[tb]
\includegraphics[scale=1]{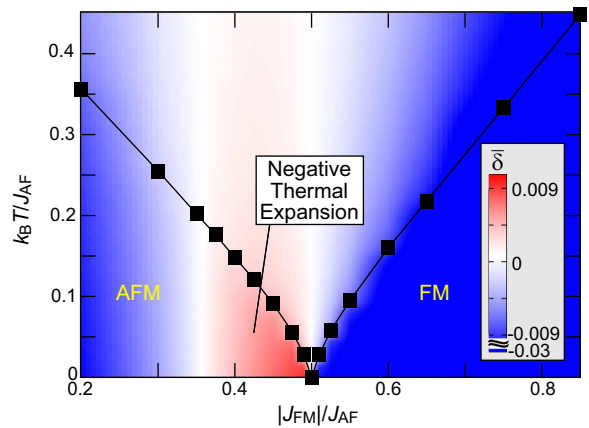}
\caption{(Color online) Phase diagram of the spin-lattice model in Eq.~(\ref{eq:model}) and color map of the averaged difference of normalized bond length $\overline{\delta}$ in the plane of temperature $k_{\rm B}T/J_{\rm AF}$ and the ratio $|J_{\rm FM}|/J_{\rm AF}$. The negative thermal expansion manifested by positive $\overline{\delta}$ is observed in an area inside the antiferromagnetic (AFM) phase on the verge of the phase boundary to the ferromagnetic (FM) phase.}
\label{Fig03}
\end{figure}
Figure~\ref{Fig03} presents a calculated phase diagram of the spin-model in Eq.~(\ref{eq:model}) and a color map of the averaged difference of normalized bond length $\overline{\delta}$ in the plane of temperature $k_{\rm B}T/J_{\rm AF}$ and the ratio $|J_{\rm FM}|/J_{\rm AF}$. We first note that the system enters the antiferromagnetic (ferromagnetic) phase with lowering temperature when $|J_{\rm FM}|/J_{\rm AF}<0.5$ ($|J_{\rm FM}|/J_{\rm AF}>0.5$). In both regions, the critical temperatures decrease as the ratio $|J_{\rm FM}|/J_{\rm AF}$ approaches 0.5. Eventually, both critical temperatures are suppressed to zero at $|J_{\rm FM}|/J_{\rm AF}=0.5$, for which the $\delta$-independent term of the exchange coupling $J$ in Eq.~(\ref{eq:nnexch}), i.e., $J_{\rm AF}-2\left|J_{\rm FM}\right|$, vanishes. We next note that the negative thermal expansion characterized by a positive value of $\overline{\delta}$ appears in a limited area within the antiferromagnetic phase near the phase boundary to the ferromagnetic phase, indicating that the antiferromagnetic order does not necessarily induce the negative thermal expansion, but it occurs only when the ratio $|J_{\rm FM}|/J_{\rm AF}$ satisfies a certain condition.

\begin{figure}[tb]
\includegraphics[scale=1]{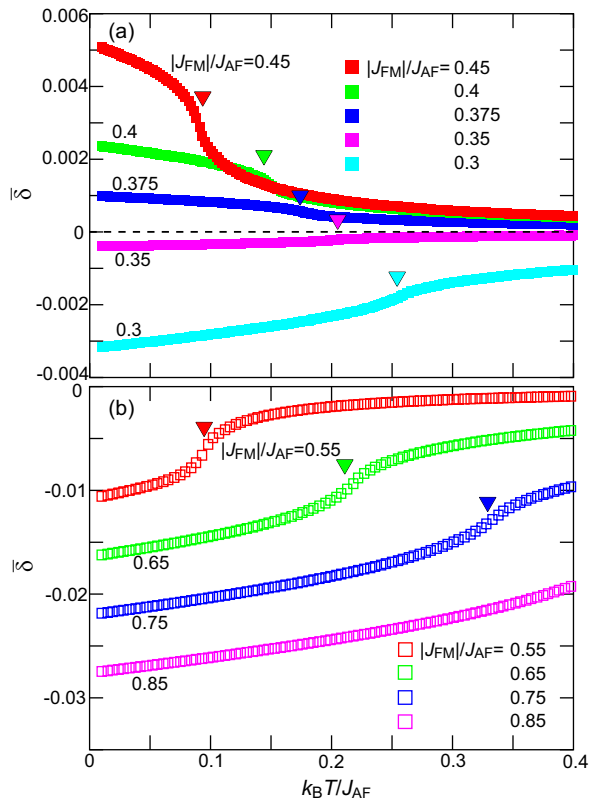}
\caption{(Color online) Calculated temperature profiles of averaged difference of the normalized bond length $\overline{\delta}$ for several values of $|J_{\rm FM}|/J_{\rm AF}$. (a) Those for $|J_{\rm FM}|/J_{\rm AF}\leq 0.45$ with which the system exhibits the antiferromagnetic order at low temperatures. (b) Those for $|J_{\rm FM}|/J_{\rm AF}\geq 0.55$ with which the system exhibits the ferromagnetic order at low temperatures. Here a positive (negative) value of $\overline{\delta}$ corresponds to the expansion (contraction) of crystal volume. The magnetic transition temperatures are indicated by inverted triangles.}
\label{Fig04}
\end{figure}
\begin{figure}[tb]
\includegraphics[scale=1]{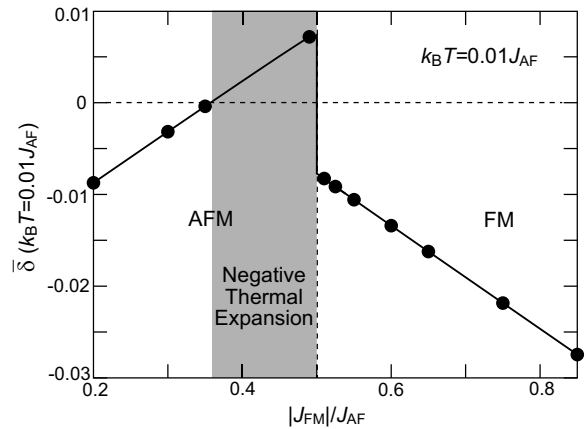}
\caption{Calculated averaged difference of the normalized bond length $\overline{\delta}$ at the lowest temperature of the Monte-Carlo simulations, i.e., $k_{\rm B}T/J_{\rm AF}=0.01$, as a function of the ratio $|J_{\rm FM}|/J_{\rm AF}$. The negative thermal expansion with $\overline{\delta}>0$ occurs in the range $0.36 \lesssim |J_{\rm FM}|/J_{\rm AF} \lesssim 0.5$.}
\label{Fig05}
\end{figure}
This aspect can be seen in the calculated temperature profiles of $\overline{\delta}$ for several values of $|J_{\rm FM}|/J_{\rm AF}$ in Figs.~\ref{Fig04}(a) and (b). Here, Fig.~\ref{Fig04}(a) shows the results for $|J_{\rm FM}|/J_{\rm AF}\leq 0.45$ where the antiferromagnetic order emerges at low temperatures, whereas Fig.~\ref{Fig04}(b) shows those for $|J_{\rm FM}|/J_{\rm AF}\geq 0.55$ where the ferromagnetic order emerges. The phase transition points are indicated by inverted triangles in both figures. The plots in Fig.~\ref{Fig04}(a) show increasing behaviors with $\overline{\delta}>0$ upon cooling when the ratio $|J_{\rm FM}|/J_{\rm AF}$ takes rather large values of 0.375, 0.4, and 0.45, and their rates of increase show noticeable rise at the phase transition points. On the contrary, we observe decreasing behaviors with $\overline{\delta}<0$ upon cooling in Fig.~\ref{Fig04}(a) when the ratio $|J_{\rm FM}|/J_{\rm AF}$ is rather small as 0.3 and 0.35. Moreover, the plots in Fig.~\ref{Fig04}(b) always show decreasing behaviors with $\overline{\delta}<0$, and they exhibit an abrupt drop at the phase transition points. These facts indicate that the negative thermal expansion occurs in the antiferromagnetic system when a rather strong ferromagnetic contribution $J_{\rm FM}$ from the 90$^\circ$ $M$-$X$-$M$ bonds competes with the antiferromagnetic contribution $J_{\rm AF}$ from the direct $M$-$M$ bond. We also find that the negative thermal expansion does not occur in the ferromagnetic system.

In Fig.~\ref{Fig05}, we plot $\overline{\delta}$ at the lowest temperature of the present simulations (i.e., $k_{\rm B}T/J_{\rm AF}=0.01$) as a function of $|J_{\rm FM}|/J_{\rm AF}$. We find that $\overline{\delta}$ is positive in the region $0.357 \lesssim |J_{\rm FM}|/J_{\rm AF} < 0.5$ indicating that the negative thermal expansion occurs when the ratio $|J_{\rm FM}|/J_{\rm AF}$ is within this region. The reason why the negative thermal expansion is observed in this limited regime can be understood by a similar argument as in our previous paper~\cite{Kobayashi19}. For the occurrence of the magnetism-induced negative thermal expansion, the following two conditions are required. First, the system should exhibit the antiferromagnetic phase transition with lowering temperature. This means that the nearest-neighbor exchange coupling $J(\delta)$ in Eq.~(\ref{eq:nnexch}) should be positive even at $\delta=0$, which gives a condition $J(0)=J_{\rm AF}-2|J_{\rm FM}|>0$. Second, the exchange coupling $J(\delta)$ in Eq.~(\ref{eq:nnexch}) should be an increasing function with respect to $\delta$ such that the crystal-volume expansion or the bond-length elongation could further stabilize the antiferromagnetic order via enhancing the antiferromagnetic exchange interaction $J$. Therefore, the coefficient of the $\delta$-linear term in Eq.~(\ref{eq:nnexch}) should be positive, which gives another condition $28|J_{\rm FM}|-10J_{\rm AF}>0$. These two conditions are summarized as $0.357 \lesssim |J_{\rm FM}|/J_{\rm AF} \leq 0.5$, and this is nothing but the required condition for which the honeycomb-lattice magnets can host the negative thermal expansion. 

It is worth mentioning that the negative thermal expansion is never triggered by the ferromagnetic ordering. This is because the following two conditions, i.e., $J(0)=J_{\rm AF}-2|J_{\rm FM}|<0$ and $28|J_{\rm FM}|-10J_{\rm AF}<0$ are required for the occurrence of ferromagnetism-induced bond elongation according to the same argument as above. These two conditions are reduced to $|J_{\rm FM}|/ J_{\rm AF}>0.5$ and $|J_{\rm FM}|/ J_{\rm AF}<0.357$, respectively, and there is no parameter range which simultaneously satisfies these conditions. We also note that these arguments do not provide any information of behaviors and properties of the negative thermal expansion and the associated magnetic phase transitions at finite temperatures. Thus, we have performed the Monte-Carlo analyses, which successfully demonstrated the thermal evolutions of bond lengths and spin correlations as well as their correlations.


\section{Conclusion and Discussion}
To summarize, we have theoretically predicted possible magnetism-driven negative thermal expansion in honeycomb-lattice antiferromagnets with networks of edge-sharing $MX_6$ octahedra, in which the nearest-neighbor exchange couplings are composed of competing antiferromagnetic and ferromagnetic contributions originating respectively from the direct 180$^\circ$ $M$-$M$ bond and the 90$^\circ$ $M$-$X$-$M$ bonds. Although both contributions decrease with increasing bond length or crystal-volume expansion, the decrease of ferromagnetic contribution is more significant, and, thereby, the nearest-neighbor exchange coupling as a sum of these opposite contributions can be enhanced upon the bond elongation in contrast to our naive expectation that the exchange coupling should be reduced by the bond elongation via weakening of the orbital hybridizations. Consequently, the crystal volume can expand when the system enters the antiferromagnetic phase with lowering temperature so as to maximize the energy gain associated with the nearest-neighbor antiferromagnetic exchange coupling. We derive required conditions for the emergence of negative thermal expansion with respect to the ratio of these two contributions as $0.357<|J_{\rm FM}|/J_{\rm AF}<0.5$. 

Note that there is a long history of research that localized spin models such as classical and quantum Heisenberg models have been successfully applied to the honeycomb-lattice magnets. In this study, we, therefore, employed a classical spin-lattice model with the localized spins $\bm S_i$ and the nearest-neighbor exchange couplings among them. Here we neglect further-neighbor exchange couplings and possible magnetic interactions originating from itinerant electrons such as the Stoner-type ferromagnetic interactions, the double-exchange interactions, and the RKKY interactions. The present work provides a rather general theory for the magnetism-induced negative thermal expansion for the honeycomb-lattice magnets without considering any specific compounds by taking the typical and simple model. We may need to consider additional interactions including the above-mentioned ones when we investigate specific materials.

Our work provides us an important guiding principle to search for new materials hosting the negative thermal expansion because there are a huge number of antiferromagnetic~\cite{Shirane1959,Chemberland1970,Tsuzuki1974,Regnault1980,Rogado2002,Tsirlin2010,Pascual2012,LeeS2012,LeeS2014,YanYJ2012,Iakovleva2019,Rani2013,Seibel2013,Tang2014,Rao2014,Rani2015,Itoh2015,McNally2015,KimSW2016,Bera2017,Sugawara2017,Haraguchi2018,Haraguchi2019,Nalbandyan2019,Tursun2019,NiM2020,Sala2021,TangYS2021,Ishii2021,Smirnova2009} and ferrimagnetic~\cite{Nakayama2011,WangY2015,KimSW2016b,WangW2021} honeycomb-lattice magnets with edge-sharing octahedra. One promising candidate is a solid solution system of the ilmenite-type manganese oxide Mg$_{1-x}$Zn$_x$MnO$_3$, in which we can change the nearest-neighbor exchange coupling from antiferromagnetic (MgMnO$_3$) to ferromagnetic (ZnMnO$_3$) by cation substitution~\cite{Haraguchi2019}. Because the negative thermal expansion is expected to occur in the antiferromagnetic phase near the boundary to the ferromagnetic phase, the solid solution system is promising because the strength ratio $|J_{\rm FM}|/J_{\rm AF}$ can be tuned rather continuously by varying the Zn concentration. In addition to the honeycomb-lattice antiferromagnets, magnets having edge-sharing $MX_6$ octahedra also provide candidate materials~\cite{ZhangB2012,Cao2015,Zhang2015,Kumada2015,Zhou2016,Cho2017,Otsuka2018}. We expect that the present work will contribute to diversify the family of materials hosting the magnetism-induced negative thermal expansion.

\section{Acknowledgment}
This work is supported by JSPS KAKENHI (Grants No. 19K21858 and No. 16H06345) and the Waseda University Grant for Special Research Projects (Project No. 2019C-253).\\

\noindent
{\it Note added in proof}: After completion of this work, we have realized Ref.~\cite{Martin12} which reported an experimental observation of theantiferromagnetism-induced expansion of the in-plane lattice constant in a honeycomb-lattice antiferromagnetBaCo$_2$(AsO$_4$)$_2$. We expect that this compound is a precious example of materials hosting the magnetism-drivennegative thermal expansion caused by the proposed mechanism. Further theoretical studies based on the first-principles calculations may prove this naive expectation in the future.

\end{document}